\documentclass{ws-procs975x65}
\usepackage{psfrag}

\def\mbf#1{\ensuremath{\mathchoice{\mbox{\boldmath$\displaystyle#1$}}
{\mbox{\boldmath$\textstyle#1$}}
{\mbox{\boldmath$\scriptstyle#1$}}
{\mbox{\boldmath$\scriptscriptstyle#1$}}}}

\def\F{\mathcal{F}}
\def\X{\mathrm{X}}
\def\detVec#1{\mbf{#1}}
\def\S{\mathcal{S}}
\def\Sinv{\S^{-1}}
\def\av#1{{\langle #1 \rangle}}
\def\avS#1{{\av{#1}_{\!S}}}
\def\N{\mathcal{N}}
\def\Np{\N_p}
\def\C{\mathcal{C}}
\def\Cp{\C_p}
\def\Xhat{\hat{\X}}
\def\max{\mathrm{max}}
\def\Hz{\mathrm{Hz}}

\def\gain{\gamma}
\begin{document}

\title{Search for Continuous Gravitational Waves: simple criterion for
  optimal detector networks} 

\author{Reinhard Prix}

\address{Max-Planck-Institut f\"ur Gravitationsphysik,
  Albert-Einstein-Institut,\\
  D-14476 Golm, Germany\\
%%  \email{Reinhard.Prix@aei.mpg.de}
%%  \texttt{LIGO-P070004-00-Z}
}

\begin{abstract}
We derive a simple algebraic criterion to select the optimal detector
network for a coherent wide parameter-space (all-sky) search for
continuous gravitational waves. Optimality in this context is defined
as providing the highest (average) sensitivity per computing cost. 
This criterion is a direct consequence of the properties of the
multi-detector $\F$-statistic metric, which has been derived recently. 
Interestingly, the choice of the optimal network only depends on the
noise-levels and duty-cycles of the respective detectors, and not on
the available computing power.
\end{abstract}

\bodymatter

\section{Multi-detector matched filtering}
\label{sec:multi-detector-f}

The $\F$-statistic\cite{jks98:_data} is a coherent matched-filtering detection
statistic for continuous gravitational waves (GWs). We follow the
expressions and notation of our previous work\cite{prix06:_searc}
(Paper~I) on the multi-detector $\F$-statistic metric. 
We consider a set of $N$ detectors with (uncorrelated) noise
power-spectra $S_\X$, where $X$ is the detector index, 
$\X = 1, ..., N$. 
Let $T$ be the total observation time spanned by the data to be analyzed. 
The corresponding multi-detector scalar product for narrow-band
continuous waves can be written as 
\begin{equation}
  \label{eq:1}
  \left(\detVec{x}|\detVec{y}\right) = T \Sinv \, \avS{x\,y}\,,
\end{equation}
where boldface notation denotes multi-detector vectors, i.e.\
$\left\{\detVec{x}(t)\right\}^\X = x^\X(t)$. 
We can allow for the fact that each detector will be in lock
only for a duration $T_\X \le T$, so each detector can be
characterized by a ``duty cycle'', $d_\X \equiv T_\X / T \le 1$.  
This is a slight, but straightforward generalization with respect to
Paper~I, and the corresponding noise-weighted time average $\avS{.}$ is defined as
\begin{equation}
  \label{eq:2}
  \avS{Q} \equiv \frac{1}{T}\,\sum_\X w_\X \int_0^T Q^\X(t)\, d t\,,
\end{equation}
where the weights $w_\X$ and the total inverse noise-power $\Sinv$ are
defined as
\begin{equation}
  \label{eq:3}
  w_\X \equiv d_\X \, \frac{S_\X^{-1}}{\Sinv}\,,\quad\textrm{where}\quad
  \Sinv \equiv \sum_{\X=1}^N d_\X\, S_\X^{-1}\,.
\end{equation}
The importance of Eq.~(\ref{eq:1}) is that it separates out the
\emph{scaling} with the total observation time $T$ and the set of
detectors (via $\S^{-1}$), from the averaged contribution $\avS{x\,y}$,
which does not scale with $T$ or the number of detectors. 
%%The scaling of $\Sinv$ with the number of detectors $N$ is easily seen
%%in the special case of equal-noise equal-duty cycle detectors, i.e.\
%%$S_\X = S_0$, and $d_\X = d_0$, in which case we  have $\Sinv = N
%%\,d_0\,S_0^{-1}$.  
In terms of this scalar product (\ref{eq:1}), the optimal
signal-to-noise ratio (SNR) for a perfectly-matched signal
$\detVec{s}(t)$ can be obtained as  
\begin{equation}
  \label{eq:4}
  \rho(0) = \sqrt{ \left(\detVec{s}|\detVec{s}\right)} = \sqrt{T\,\Sinv}\,\sqrt{\avS{s^2}}\,.
\end{equation}
%%where here we are only interested in the \emph{scaling} with $T$ and
%%the set of detectors $X$.
It is obvious from this expression that the SNR increases when
increasing the observation time $T$ or the number of detectors $N$.
%%For small parameter-space searches that are not computationally
%%limited, such as \emph{targeted} searches, increasing the number of
%%detectors therefore \emph{always} increases the sensitivity of the
%%search. 
%%The question of the ``optimal network'' is therefore trivial,
%%as one would always use as many detectors as possible, provided the
%%corresponding gain in SNR is still worthwhile.
However, here we are interested in the case of \emph{wide parameter-space} 
searches, in which the highest achievable SNR is computationally limited.
We therefore need to find the optimal sensitivity \emph{per computing cost}.
%%, and the conclusion about the optimal network differs from the
%%case where computing cost is not an issue.

\section{Optimizing sensitivity per computing cost}
\label{sec:optim-sens-per}

For simplicity we only consider the sensitivity to an ``average''
sky-position, so we disregard the dependence of $\avS{s^2}$ to both
the sky-position as well as the relative orientation of the different
detectors. Both should be small effects on average.
In addition to Eq.~(\ref{eq:4}) for the SNR, the second ingredient 
for the optimal network is the computing cost of a wide-parameter search.
For the sake of example we consider a search for GWs from unknown
isolated neutron stars, with unknown intrinsic GW frequency $f$,
sky-position $\alpha, \delta$ and one spindown-parameter
$\dot{f}$. One can show\cite{prix06:_searc} that in this case the
number of required templates $\Np$ scales (at least) as $\Np \propto T^6$, 
which severely limits the computationally affordable observation time $T_\max$. 
Most importantly, however, the number of templates does \emph{not}
scale with the number $N$ of detectors\cite{prix06:_searc}.
The corresponding computing cost $\Cp$ required to search these $\Np$
templates can be estimated as $\Cp \propto N\,T^7$ for a
``straightforward'' computation, while it could be reduced down to
about $\Cp\propto N\,T^6$ if the FFT-algorithm is used\cite{jks98:_data}. 
Generally, we can write
\begin{equation}
  \label{eq:6}
  \Cp \propto N \, T^\kappa\,,
\end{equation}
where typically $\kappa \sim 6 - 7$ for isolated neutron-star searches.
The linear scaling with $N$ comes from the fact that we need to
compute the correlation of each template with each of the $N$
detector time-series $x^\X(t)$. 

The question we are trying to answer is the following: for given
computing power $\Cp$ and a set of $N$ detectors, which (sub)-set of
$\hat{N}\le N$ detectors $\{\Xhat\} \subseteq \{\X\}$ yields the
highest SNR? 
Using (\ref{eq:6}), we can express $T_\max \propto \left(\Cp / N\right)^{1/\kappa}$, 
and inserting this into (\ref{eq:4}), we find 
$\rho(0) \propto \Cp^{1/(2\kappa)}\, \sqrt{ \gain(\{\X\}) }$, where 
the ``gain function'' $\gain$ is defined as
\begin{equation}
  \label{eq:8}
  \gain(\{\X\}) \equiv N^{-1/\kappa}\, \sum_{\X = 1}^{N} d_\X \,S_\X^{-1}\,.
\end{equation}
This simple algebraic function provides the sought-for criterion for
the optimal detector-network $\{\hat{\X}\}$, depending only on the
respective noise-floors $S_\X$ and duty-cycles $d_\X$. The optimal
detector network is simply the subset $\{\Xhat\}$ of detectors that
maximizes the gain-function $\gain(\{\Xhat\})$.

This optimal subset can be found in the following simple way: we label 
the detectors $X$ in order of \emph{decreasing} $d_\X S_\X^{-1}$, and
include exactly the first  $\Xhat = 1, ..., \hat{N}$ detectors in
(\ref{eq:8}) where $\gain$ reaches a maximum. It is easy to see that
this arrangement is optimal, as either adding further detectors, or
replacing any term $d_{\hat{\X}} S_{\hat{\X}}^{-1}$ in the sum by
another detector  $\X' > \hat{N}$ reduces $\gain$.   

In the special case of identical detectors, the gain
function $\gain$ is strictly monotonic with $N$, and so the optimal
network simply consists of using as many detectors as possible,
reducing the observation time $T$. 
%%This is an obvious consequence of
%%the fact that increasing $T$ is much more costly than increasing the
%%number of detectors (see Eq.~(\ref{eq:6})), while yielding the same
%%gain in SNR.     

\section{Example application}
\label{sec:example-application}

As an example, consider a set of ``typical'' detectors as given in
Table~\ref{tab:detectors}. The assumed parameters are: LIGO (H1, H2,
L1) at design sensitivity, with S5 duty-cycles, GEO (G1) at S5 sensitivity,
and S4 duty-cycle, Virgo (V2) at design sensitivity, assuming a ``typical''
LIGO duty-cycle. 
\begin{table}[htbp]
  \centering
  \tbl{Example set of detectors with ``typical'' sensitivities and duty-cycles.} 
  {
    \begin{tabular}[htbp]{ c | c || c | c | c | c | c}
     &  Frequency                                 &  H1      & L1      &  H2     & G1      &    V2       \\\hline\hline
 $d_\X$              &     ---                    &  0.71    & 0.59    &  0.78   & 0.97    &    0.7       \\
$\sqrt{S_\X}~[10^{-23}/\sqrt{\Hz}]$ &  $f = 200~\Hz$ & 2.9     & 2.9     &  5.8    & 73      &    4.4       \\
$\sqrt{S_\X}~[10^{-23}/\sqrt{\Hz}]$ &  $f = 600~\Hz$ &  7.5     & 7.5     &  15     & 39      &    5.5	   \\
     \end{tabular}
     \label{tab:detectors}
   }
\end{table}
\begin{figure}[htbp]
  \hspace*{-0.5cm}
  \psfrag{sqrt(g)}{$\sqrt{\gain}$}
  \mbox{\includegraphics[clip,width=0.55\textwidth]{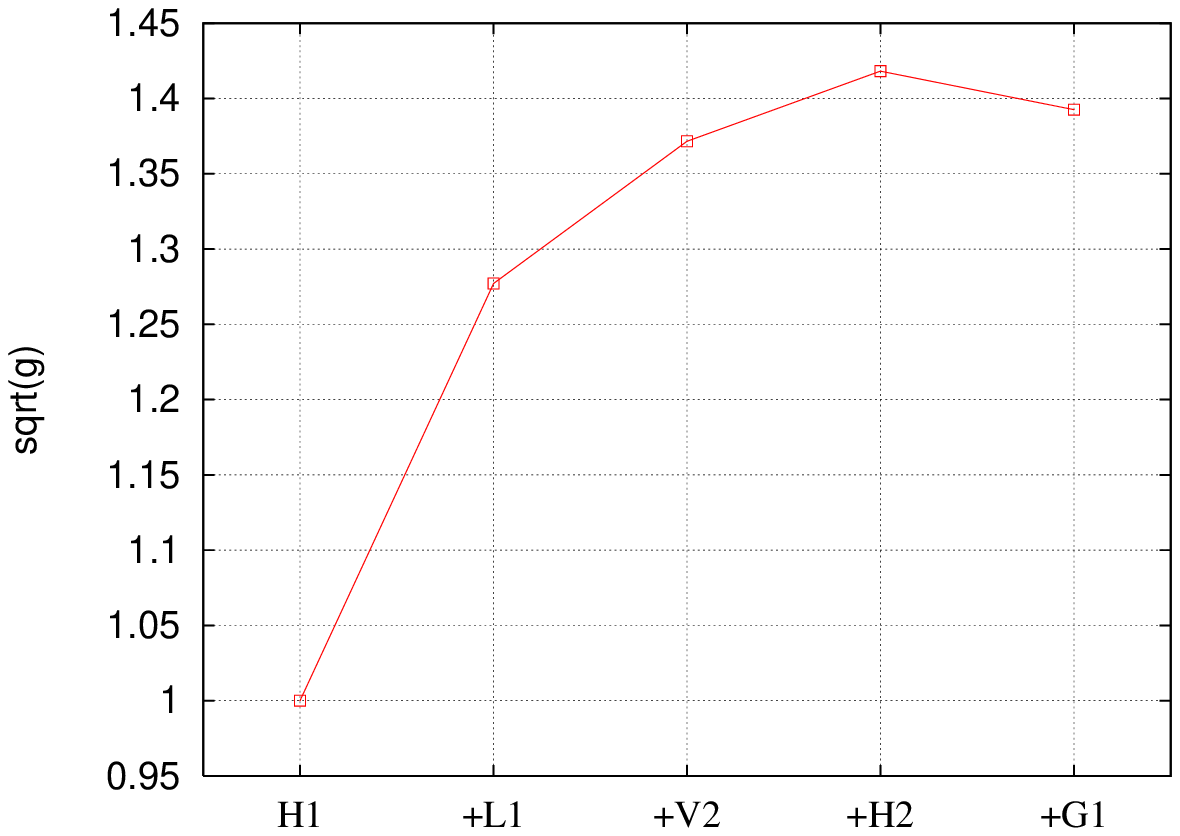}
    \hspace*{-0.5cm}
  \includegraphics[clip,width=0.55\textwidth]{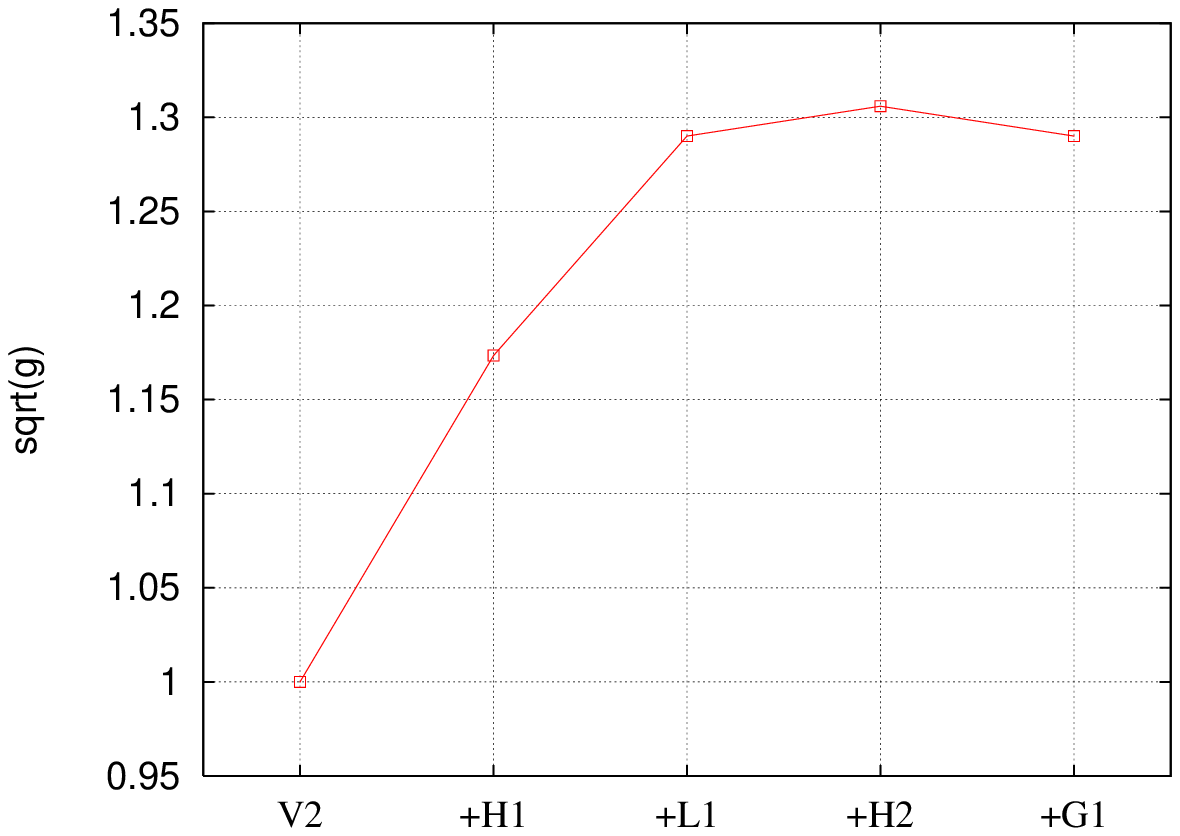}}
  \caption{SNR gain $\sqrt{\gain(\{X\})}$  (assuming $\kappa = 6$) as
    a function of the detector network, normalized to the single-detector case
    \textit{Left figure:} at f = 200~Hz. \textit{Right figure:} at f = 600~Hz.} 
  \label{fig:networks}
\end{figure}
We see that our simple criterion tells us that for a search at $f =
200~\Hz$ we should include H1, L1, V2, and H2 for the best all-sky
sensitivity per computing cost, gaining on average a total of about
$40\%$ in SNR over H1 alone. Similarly, at  $f = 600~\Hz$, we find the
same set of detectors to be optimal, with H2 providing a smaller
marginal improvement. 

\bibliography{biblio}

\end{document}